\documentclass[lettersize,journal]{IEEEtran}
\usepackage{amsmath,amsfonts}
\usepackage{algorithmic}
\usepackage{algorithm}
\usepackage{array}
\usepackage[caption=false,font=normalsize,labelfont=sf,textfont=sf]{subfig}
\usepackage{textcomp}
\usepackage{stfloats}
\usepackage{url}
\usepackage{verbatim}
\usepackage{graphicx}
\usepackage{booktabs}
\usepackage{callouts-box}
\usepackage{tabularx}
\usepackage{enumitem}
\usepackage{hyperref}
\usepackage[style=numeric, sorting=none]{biblatex}
\newcolumntype{Y}{>{\centering\arraybackslash}X}

\newcommand{\celllist}[1]{%
	\begin{minipage}[t]{0.78\linewidth}
		\vspace{0pt}
		\raggedright
		\begin{itemize}[leftmargin=1.2em,itemsep=0pt,parsep=0pt,topsep=0pt,partopsep=0pt]
			#1
		\end{itemize}
	\end{minipage}%
}

\begin{document}

\title{AI Assurance in UK Defence:\\ Challenges in Operationalising  JSP 936}

\author{Callum Cockburn,~\IEEEmembership{Senior Technical Innovation Manager,~Synoptix,}\\
		Sam Farrow,~\IEEEmembership{Technology Director,~Synoptix}
        % <-this % stops a space
\thanks{This paper was originally published in December 2025 by Synoptix as an industry whitepaper. This version extends this writing to incorporate full referencing. Whilst the majority of the text has stayed consistent, there are some minor wording changes.}% <-this % stops a space
}

\markboth{8th June ~2025}%
{}

\IEEEpubid{Approved External \copyright~2026 Synoptix}

\maketitle

\begin{abstract}
This report examines practical challenges in operationalising JSP 936 Part 1 for AI assurance in UK Defence. Using a structured interpretive review of the directive’s requirements, the analysis identifies eight thematic challenge areas: adequacy of evidence and argument, management of human interaction with AI, definition of the operational environment, integration of AI within systems of systems, assessment and maintenance of AI performance, analysis of safety and security, measurement of ethicality, and mitigation of the inherent complexities of AI. The report argues that JSP 936 provides a useful governance basis, but that implementation depends on unresolved technical, organisational, and assurance questions. These challenges stem from the socio-technical nature of AI-enabled systems, uncertainty in real-world deployment contexts, limitations in current assurance methodologies, and tensions between performance, safety, human oversight, security, and ethical acceptability. The report identifies areas where further methods, guidance, and organisational capability are needed for the ambitious, safe, and responsible adoption of AI across Defence. This is consistent with MOD’s own framing of JSP 936 as requiring iterative implementation and supporting guidance
\end{abstract}


\section{Introduction}
\IEEEPARstart{T}{his} report analyses selected requirements drawn from JSP 936 Part 1, Version 1 (“JSP 936”) \cite{ministry_of_defence_jsp_2024}. These requirements are grouped into eight thematic areas, each reflecting a recurring implementation problem. For each area, the report identifies assurance questions and discusses their implications.
The analysis does not challenge the purpose or structure of JSP 936. The principal difficulty lies in translating its requirements into technical, operational, and organisational practice. Operationalisation presents complex and interdependent challenges. This report identifies and analyses these challenges, to inform further methodological development and implementation guidance, as well as support wider awareness of how these challenges may impact the Ambitious, Safe, and Responsible deployment of AI across UK Defence.

\subsection{Development Process}
A manual extraction of requirements from JSP 936 Part 1 revealed 272 requirements (121 “should”, 151 “must”) that make up this directive. Some statements would require further decomposition to meet requirements quality standards, so the final number of atomic requirements would likely be higher, but this serves as an initial benchmark.
A manual review was undertaken of the requirements, where each requirement was categorised by difficulty. Approximately 35 were identified in the “significant” category: those that represent a significant challenge with current maturity of AI technology or techniques for undertaking AI assurance. These were clustered into 8 key challenges by identifying these and trends, which are represented in Table \ref{tab:challenges}. Whilst this is a subjective interpretation, it does meet this report’s intended purpose. It is not intended to be a complete record of every challenge with implementing JSP 936, simply to highlight some key issues that this might bring.
This analysis was purely based on JSP 936 Part 1, version 1. It is important to note that JSP 936 Part 2 and the AI Practitioner’s Handbook may provide best practice, guidance, and methods to mitigate some of these challenges. 
\IEEEpubidadjcol

\begin{table*}[]
	\centering
	\caption{Our analysis highlights 8 key challenges for implementing JSP 936.}
	\label{tab:challenges}
	\normalsize
	\begin{tabularx}{\textwidth}{Y Y Y Y}
		\toprule
		\multicolumn{4}{c}{\textbf{Challenges with Implementing JSP 936}}                                                                                                                                     \\
		\midrule
		Justifying Adequacy of Evidence and Argument & Managing Human Interaction with AI                  & Defining the Operational Environment & Dealing with AI as a System of Systems Component \\
		Assessing and Maintaining AI Performance     & Analysing Safety and Security in AI-enabled systems & Measuring Ethicality                 & Mitigating the Inherent Complexities of AI\\
		\bottomrule      
	\end{tabularx}%
\end{table*}

\subsection{Challenges}
For each challenge, the relevant requirements from JSP 936 are identified (including a link to the Directive paragraph in question), and a series of questions and sub-questions are posed. Where relevant, there is a brief discussion around some of the issues or challenges contained within each question.

\section{Justifying Adequacy of Evidence and Argument}
	\begin{notebox}[Requirements from JSP 936]
		“Risk Owners must judge that the evidence supporting confidence in the system is adequate throughout the AI lifecycle” (P. 6)\\
		“Data should be demonstrated as correct” (P.157)\\
		“Where High-Level Requirement behaviours are to be implemented through AI and are not directly decomposable into Low-Level Requirement, the combination of the training algorithm and data requirements must be demonstrated as meeting the intent of the High-Level Requirement” (P.144)
	\end{notebox}

\subsection{Judgement of Adequacy}
\begin{quote}
	\textbf{\textit{	How do we judge that evidence is adequate?\\
	Does adequacy change in different contexts?}}
\end{quote}
Assurance is commonly used as a mechanism to demonstrate compliance with requirements, to support acceptance by stakeholders that their requirements have been met, or to support certification against applicable regulation of standards \cite{bloomfield_safety_2010, rhodes_software_2010, habli_big_2025}. Adequacy, therefore, supports the demonstration of justified trust, where those accountable for systems are attempting to demonstrate evidence that they have sufficiently addressed information asymmetries by measuring, evaluating, and communicating reliable evidence \cite{gazendam_mind_2023, department_for_science_innovation_and_technology_introduction_2024}. For AI systems, where they operate in a complex socio-technical space \cite{kudina_sociotechnical_2024}, this means that there are a significant number of dimensions of adequacy, including:
\begin{itemize}
	\item \textbf{Legal adequacy}: do we have a robust and legally defensible argument against possible and reasonable civil and criminal action? Are we content that the remaining legal risk has been suitably mitigated? Have we met our specific legal obligations (e.g. Article 36 review justifications)?
	\item \textbf{Ethical adequacy}: are we operating within the general ethical consensus of civil society, and the framework of reasonable ethical consideration - bearing in mind that ethical acceptability is highly subjective and variable? As well as operating ethically, are we are being seen to be ethical? Would a reasonable outside observer be able to sit in on our assurance review meetings and be content that we’re considering and thinking about ethical issues as part of our development?
	\item \textbf{Technical and safety adequacy}: do we have sufficient evidence that the system (in the real world) performs as expected? Do we have sufficient evidence that the hazards possible within the system are well controlled?
\end{itemize}

These cannot be considered in isolation from each other - a technical inadequacy could lead to an ethical inadequacy which could then lead to a legal inadequacy, for example. They also can’t purely consider an AI model - they need to integrate perspectives around a sociotechnical AI system deployment.

\subsection{Adequacy of Data}
\begin{quote}
	\textbf{\textit{How can data be “correct” – given that data representing a real-world system will only ever be an abstraction?}}
\end{quote}

Beyond a small set of trivial real-world problems, it is unattainable for data about a real-world system to be treated as a literally "correct" representation of reality. Data are produced through processes of selection, measurement, abstraction, and interpretation, and data models should therefore be judged by their adequacy for a defined purpose rather than by comparison with an ideal of perfect representation  \cite{lever_facing_2025, bokulich_data_2021}. The question is to define the sufficiency of this representation, such that adequate coverage of the features, conditions, and variations that are relevant to system performance within its intended operational design domain are achieved.

This is especially true in the situations which AI systems are often proposed for us - where the real world is complex and uncertain \cite{defence_committee_developing_2025}.  Rather than being correct, what we really require is that:
\begin{quote}
\textit{The dataset should provide sufficient coverage of the features, conditions, and sources of variation that are relevant to the system’s intended use and operational design domain, such that the trained model can generalise appropriately under expected conditions of deployment.}
\end{quote}

This is still challenging and a significant effort, but more accurately describes the sufficiency challenge presented as part of the non-functional requirements of the system \cite{habibullah_non-functional_2023}. Whilst JSP 936 does recognise the criteria defined in Ashmore et al. \cite{ashmore_assuring_2022} (relevance, completeness, balance, and accuracy), the correctness framing may indeed be more hindrance than helpful, in this context. 
\subsection{Requirements for AI}
\begin{quote}
	\textbf{\textit{How do we map requirements to AI system design?\\
	How do we determine adequacy against “intent” of high-level requirements?}}
\end{quote}

 In many ML product-development contexts, AI systems, and especially machine-learning-based systems, are not developed from a stable set of formal system requirements in the way expected in conventional systems engineering. In practice, development is often organised through Agile or Agile-like workflows, with work framed through user stories, experiments, and backlog items rather than a complete set of verifiable system-level requirements \cite{romao_agile_2026, amershi_software_2019}. This is workable for many software products, but it creates difficulties in conventional acquisition settings, where traceability, verification, and contractual clarity are central. A major reason is methodological: conventional systems engineering is largely deductive, beginning with explicit requirements and allocating them to system elements, whereas many machine-learning components are often developed inductively, with system behaviour inferred from training data rather than fully derived from ex ante specification \cite{habiba_how_2024}. These difficulties and trade-offs become more pronounced when AI is embedded within a larger system and across a full system lifecycle, particularly where hardware, operational, safety, or interface constraints must also be specified and assured \cite{ahmad_requirements_2023, ashmore_assuring_2022, nalchigar_modeling_2021}.
Requirements engineering for AI is also not uniform across application types. The requirements relevant to a computer vision system differ materially from those relevant to a natural language processing system, for example. A computer-vision system may require requirements for image resolution, frame rate, lighting conditions, sensor placement, and environmental variation. A natural-language-processing system may instead require requirements for language coverage, dialect, context sensitivity, semantic interpretation, response latency, and privacy. Whilst this list is not exhaustive, it demonstrates the variance in needs from different types of AI systems, and the corresponding difficulty in defining consistent requirements. 

In a general sense, some recurrent and important challenges in applying conventional requirements engineering to machine learning include:

\textbf{Stakeholder Expectations:}
\begin{itemize}
	\item Stakeholders often have high expectations for AI systems but struggle to understand their true capabilities and limitations, which can lead to unrealistic demands and misaligned goals \cite{alves_status_2023, kinney_expectation_2024}.
\end{itemize}	
\textbf{Defining Requirements:}
\begin{itemize}
	\item Writing precise specifications for data-driven features is challenging, especially when concepts like “a person” need machine-comprehensible definitions \cite{ahmad_requirements_2023}.
	\item Requirements are difficult to draft when the necessary data is not yet available, creating uncertainty in early stages.
	\item Incorporating diverse considerations (e.g. legal, ethical) add further complexity to defining requirements for AI systems \cite{nalchigar_modeling_2021}.
	\item Many non-functional requirements (NFRs) are harder to define for AI systems and can be challenging to verify, particularly because ML-related qualities may differ in definition, measurement, scope, and relative importance from conventional software NFRs \cite{habibullah_non-functional_2023}. In addition, these requirements can easily lead to optimization for the wrong objectives, due to the strong influence on system performance that training-time parameter and design choices can have. This means that ensuring alignment with intended outcomes is a persistent challenge.
\end{itemize}
\textbf{Nature of AI:}
\begin{itemize}
	\item Many AI systems are probabilistic or non-deterministic, making exhaustive testing and verification difficult, especially in safety-critical settings where assurance evidence is required across the ML lifecycle \cite{habibullah_non-functional_2023, ashmore_assuring_2022}.
	\item Emergent behaviours cannot be fully anticipated or specified in advance, especially for complex, adaptive, or multi-agent systems, adding unpredictability to system design.
\end{itemize}	
\textbf{Trade-offs and Optimisation Factors:}
\begin{itemize}
	\item Balancing competing non-functional requirements is difficult because ML qualities may be defined and measured over the model, data, or whole system, and design work may involve explicit trade-offs among data, algorithms, and intended outcomes \cite{habibullah_non-functional_2023, nalchigar_modeling_2021}.
\end{itemize}
\textbf{Skillsets and Experiences:}
\begin{itemize}
	\item Data scientists or AI engineers typically lack deep expertise in formal requirements engineering, while requirements engineers often have limited experience with AI systems \cite{amershi_software_2019, habiba_how_2024}.
	\item The lack of and low maturity of established guidelines in this area compounds the difficulty, leaving teams without clear best practices.
\end{itemize}

\section{Managing Human Interaction with AI}
\begin{notebox}[Requirements from JSP 936]
	“The fact that higher levels of autonomy typically reduce the potential for human decision-making within the control loop must be considered” (P.35)\\
	“An assessment must be made of the impact that AI could have on the identified human stakeholder groups” (P.52)\\
	“An analysis of the allocation of functions between human and AI agents and AI behaviours across all modes of function and levels of autonomy must be conducted” (P.125)\\
	“Training Needs Analysis for users of AI-based systems should include consideration of the users’ need to develop an understanding of system behaviour, performance and limitations and calibrate their trust in the system under different use cases and conditions” (P.130)\\
	“In cases where AI is supporting important/risky decision-making, its output should have additional checks applied that are undertaken by a relevant subject matter expert” (P.180)
\end{notebox}
\subsection{Levels of Autonomy}
\begin{quote}
	\textbf{\textit{How does higher levels of autonomy interact with other requirements? Does it “raise the bar” to consider adequacy of confidence?\\
	How do you manage defining interactions at different levels of autonomy? Which ones are common between different levels, and which ones change?}}
\end{quote}
The first question, prior to discussing the level of assurance required at each autonomy level, is defining the levels themselves. Principal dimensions include decision-making authority, execution of actions, information monitoring, and handling of exceptions or fallback scenarios. These dimensions range from full human control to system autonomy across specified functions, rather than necessarily applying uniformly to the whole system. Some examples of levels include:
\begin{itemize}
	\item Lloyd’s Register outlines a progression from manual control (AL0) to un-supervised autonomy (AL6), emphasising increasing independence in decision/action, and where AL6 denotes a fully autonomous ship with no access required during a mission \cite{lloyds_register_lr_2017}.
	\item SAE J3016 distinguishes automation levels by the roles of the human user and the driving automation system in performing the dynamic driving task and fallback performance \cite{society_of_automotive_engineers_taxonomy_2024}.
	\item Parasuraman et al. propose a model of types and levels of automation, spanning information acquisition, information analysis, decision/action selection, and action implementation, with automation possible at different levels in each function  \cite{parasuraman_model_2000}.
	\item ISO 23860  introduces a matrix of control and automation degrees, identifying configurations such as Fully Autonomous (FA), Autonomous Control (AC), Operator-Automation (OA), and Operator Exclusive (OE), based on the balance of human and system responsibilities \cite{international_standards_organisation_ships_2022}.
\end{itemize}
Together, these frameworks highlight autonomy as a multidimensional construct involving control distribution, system capability, function allocation, fallback responsibility, and operational context. They also show that autonomy levels are not defined consistently across domains, which complicates assurance arguments that depend on the meaning of a claimed autonomy level.

The level of assurance required for autonomy depends on multiple interrelated factors that influence safety, reliability, and human-system interaction. These include the system’s capacity to affect the environment, the degree and effectiveness of human monitoring, and the ability to intervene during error states. This is particularly important where assurance depends on human monitoring or intervention, because operators of automated systems may be less able to take over after automation failure due to reduced situational awareness, passive monitoring, or skill degradation \cite{bainbridge_ironies_1983, endsley_out-loop_1995}.

Assurance must account for the risk inherent in the operational context, the type of task, and the automation’s role across the decision chain, including information acquisition, information analysis, decision/action selection, and action implementation \cite{parasuraman_model_2000}. Human performance considerations - such as mental workload, situational awareness, complacency, skill degradation, and trust – directly impact assurance needs \cite{endsley_out-loop_1995, parasuraman_complacency_2010}, as do measures of automation performance and reliability, and the ability to communicate system purpose, process, and performance in a way that supports appropriate reliance \cite{lee_trust_2004}. Finally, the consequences and costs of incorrect decisions or actions determine the rigor of verification, validation, and monitoring required to certify autonomy at a given level.

Reusing assurance evidence for autonomy is challenging because assurance claims are tied to assumptions about the system, its operating context, and the allocation of responsibilities between humans and automation. When those assumptions change, prior evidence may no longer support the same safety claim \cite{hawkins_guidance_2022, asaadi_dynamic_2020}. Managing mixed-criticality architectures, or architectures that combine conventional components, autonomy functions, and human oversight, adds complexity, requiring clear segregation and assurance strategies \cite{hawkins_guidance_2022}. Furthermore, the definitions, assumptions, and limitations underlying previous assurance arguments must be explicitly articulated, because safety cases depend on structured claims, supporting evidence, and the operating context in which those claims are valid \cite{bloomfield_assurance_2021}. Relevance assessment is also essential - evidence must be validated against the new operational context and autonomy level to confirm that prior arguments remain applicable and do not introduce hidden risks. 

In addition to validation at different levels of autonomy, reversionary modes require explicit analysis. Validation is needed not only for operation within each mode, but also for entry into and exit from that mode - the state transition itself \cite{yang_framework_2024}. This is required because operators may be poorly placed to resume control after automation failure if automation has reduced their situation awareness or shifted them into passive monitoring \cite{endsley_out-loop_1995}.

\subsection{Impact Assessments}
\begin{quote}
	\textbf{\textit{How do you avoid impact assessments becoming cumbersome “tickbox” exercises?}}
\end{quote}
The literature on AI impact assessments (AI-IAs) is growing, but there remains limited settled evidence on how AI-IAs should be structured, implemented, and evaluated in practice. Stahl et al.'s systematic review found some convergence across AI-IA approaches, but concluded that the field has not reached agreement on content, structure, or implementation \cite{stahl_systematic_2023}. Although they are mandated in many AI management standards and regulation (including ISO 42001 \cite{international_standards_organisation_isoiec_2023}, EU AI Act \cite{european_parliment_eu_2023}), research identifies difficulties in conducting them in ways that reliably identify and mitigate unwanted ethical and societal harms. Stahl et al. frame AI-IAs as a way to identify both positive and negative impacts early, which is important because AI systems are normally deployed precisely to change decisions, processes, or outcomes \cite{stahl_systematic_2023}.

For an AI-IA (or indeed, other similar such assessments) to avoid becoming a purely procedural exercise, the assessment needs to be connected to measurable impacts, trade-offs, affected stakeholder groups, update triggers, and decision points \cite{methnani_let_2021, stahl_systematic_2023}. The following are therefore author-synthesised requirements for making the assessment operationally useful:
\begin{itemize}
	\item define and quantify impact in ways that allow us to measurably identify what impact matters, and also that the measures of impact are those that are consequential in the outputs and outcomes created by the system.
	\item develop systematic ways to develop cost-benefit, performance-impact, risk-reward, or other trade-offs between different system architectures, designs, realisations, or implementations.
	\item manage concerns between individuals, groups of individuals, societies, organisations, governments, market forces, national states and geopolitics, and international considerations.
	\item develop ways and mechanisms for ensuring that impact assessments stay up to date with system capabilities and use cases, so that assessments stay up to date.
	\item develop ways and mechanisms for assessing the impact of “undesigned” or emergent capabilities of the system \cite{berti_emergent_2025} (particularly with more general-purpose AI systems \cite{triguero_general_2024}).
	\item define how the content of these assessments will be assured, including whether they are subject to independent scrutiny or are primarily internal assurance artefacts for organisational decision-making.
	\item understand how do we operationalise and integrate these impact assessments into the actual decisions that matter - both high-level usage and deployment decisions, but also low-level technical design and architecture decisions where significant change to the system can occur?
\end{itemize}

\subsection{Confidence and Calibration}
\begin{quote}
	\textbf{\textit{How can we define the confidence that someone has in an AI system? How do we understand the basis for this confidence – understanding which factors influence decision? \\ Is it a problem if users are “calibrated” against “wrong” factors?}}
\end{quote}
In the context of operator-machine interactions, trust is an attitude that is relevant to automation in situations that include:
\begin{itemize}
	\item levels of uncertainty,
	\item a cooperative relationship between at least two entities,
	\item some exchange between the two entities.
\end{itemize}
This framing follows Lee and See's account of trust in automation, where trust guides reliance when complexity, uncertainty, and unanticipated situations make complete understanding of the automation impractical \cite{lee_trust_2004}.
Where there is no meaningful uncertainty or vulnerability, trust becomes less analytically important because reliance does not expose the user to material risk. Given that uncertainty is central to trust, there is also the possibility of wrongly trusting or wrongly distrusting the system. This leads to the idea of calibrated trust: the degree to which the user's trust matches the automation's true capabilities in the relevant context. 

Trust can be affected by system-related, human-related, and context-related factors. The examples in Table \ref{tab:trust_calibration} are consistent with empirical reviews that identify dispositional, situational, and learned contributors to trust in automation \cite{kaplan_trust_2023, hoff_trust_2015}.

\begin{table*}[]
	\centering
	\caption{A significant number of factors affect trust calibration.}
	\label{tab:trust_calibration}
	\normalsize
	\begin{tabularx}{\textwidth}{Y Y Y}
		\toprule
		\textbf{System-related components} & \textbf{Human-related components} & \textbf{Context-related components} \\
		\midrule
		Performance-based         & Ability-based            & Tasking-related \\
		\midrule
		\celllist{
			\item Dependability
			\item Performance
			\item Predictability
			\item Reliability
		} &
		\celllist{
			\item Competency
			\item Expectancy
			\item Expertise
			\item Prior Experience
			\item Workload
		} &
		\celllist{
			\item Risk
			\item Task complexity
			\item Task type
		}\\
		\midrule
		Attribute-based           & Characteristic-based     & Teaming-related            \\
		\midrule
		\celllist{
			\item AI Personality
			\item Anthropomorphism
			\item Appearance
			\item Behaviour
			\item Communication
			\item Level of Automation
			\item Reputation
			\item Transparency
		} &
		\celllist{
			\item Attitudes toward AI
			\item Comfort with AI
			\item Culture
			\item Education
			\item Personality traits
			\item Propensity to trust
			\item Satisfaction
		} &
		\celllist{
			\item Communication
			\item Interaction frequency
			\item Shared mental models
			\item Tenure
		} \\
		\bottomrule      
	\end{tabularx}%
\end{table*}
Trust calibration can be understood across several dimensions, although there is a considerable variance of maturity in the understanding of their development. At its simplest, calibration is the balance between justification and trust - whether the human’s perception of the machine capabilities is appropriately aligned to the true, “ground truth”, capabilities \cite{hoffman_trust_2013, bollaert_measuring_2024}. There are also additional complexities, such as:
\begin{itemize}
	\item Exogenous vs. endogenous calibration distinguishes whether trust is adjusted before/after interaction (exo) or during interaction through interventions (endo) \cite{wischnewski_measuring_2023}. 
	\item Warranted vs. unwarranted calibration addresses whether trust accurately reflects system reliability or is influenced by factors like reputation or anthropomorphism \cite{lee_trust_2004, wischnewski_measuring_2023}. 
	\item Static vs. adaptive calibration considers whether trust adjustments remain fixed or dynamically adapt to user needs and behaviour over time \cite{okamura_adaptive_2020, lebiere_adaptive_2021}. 
	\item Performance-oriented vs. process-oriented calibration differentiates between providing reliability metrics and explaining system processes; the latter requiring users to interpret how process details relate to performance \cite{marusich_trust_2025, guo_building_2026}.
\end{itemize}
All of these factors combine to result in a complex, delicate balance, which requires careful management and engineering to avoid unintended consequences; trust must be modelled as dynamic and influenced by user characteristics, situational context, learned experience, and automation display characteristics \cite{hoffman_trust_2013}.

\subsection{Expertise, Oversight, and AI}
\begin{quote}
	\textbf{\textit{If the human needs to be able to understand the system’s outputs, what information do they need to see? How can we determine which of the inputs to the system were important, and need to be provided to the user? \\ What is the minimum effective level of information that we need to provide to the operator to allow them to gain adequate situational awareness?\\How do we design effective oversight and interactions mechanisms, so that operators are able to effectively interact with and oversee systems?\\How could the human-machine oversight interaction breakdown and fail? What might cause it to fail, and what would the consequences be? How can we design systems to be resilient to oversight failures?\\What does an SME for an advanced AI system look like? Does expertise in the performance of traditional systems translate to expertise in the performance of advanced AI systems (particularly when integrated elements of autonomy)?}}
\end{quote}
\subsubsection{Transparency and Situational Awareness}
We can’t provide an overseeing operator (which could be a subject matter expert but may not be in many situations) with all information about the system, its environment, and its decisions. Providing all available information would often exceed operator cognitive capacity and defeat the purpose of effective oversight. Therefore, we have to produce a meaningful subset of information: the minimum effective information needed for the operator to maintain adequate situation awareness and intervene when required \cite{endsley_out-loop_1995, endsley_supporting_2023}.
This will be context dependent, but the questions can be aligned to the classic situation awareness levels of perception, comprehension, and projection \cite{endsley_out-loop_1995}. Combining these into 3 main questions of situational awareness \cite{chen_situation_2018}:
\begin{itemize}
	\item \textbf{What is going on, and what is the system trying to achieve?}
	\\→	Key elements: purpose, goals, process, intentions, progress, performance
	\item \textbf{Why is the system taking this action?}
	\\→	Key elements: reasoning process, belief, purpose, constraints (environmental and otherwise)
	\item \textbf{What should the operator expect to happen?}
	\\→	Key elements: potential limitations, uncertainty, likelihood, history of performance, projection to future/end state
\end{itemize}

\subsubsection{Human-Machine Interaction and Oversight}
Particularly when interacting with autonomous systems, human-machine interaction is a complex subject. Current debates often reduce these interactions to simple forms, such as human-in-the-loop, human-on-the-loop, or human-out-of-the-loop. These categories do not capture whether the human has the information, authority, time, and interaction mechanisms needed for effective oversight, nor whether the system supports shared understanding and calibrated trust, especially in complex systems or real-world environments \cite{chen_situation_2018, sterz_quest_2024}.

Oversight design should be tied to the objectives of AI use, the allocation of human and automation responsibilities, and the risk factors of the system and application \cite{hawkins_guidance_2022, sterz_quest_2024}. For example, if we are aiming to achieve speed of processing, the oversight mechanisms might look very different to if we are aiming to achieve compliance or quality objectives. Given the complex nature of these systems, it’s also likely that there are multiple system objectives that need to be balanced, or even objectives that vary over different sub-components of the system. Equally, core risk factors of the application, such as the severity of consequences for incorrect decisions, the time sensitivity and uncertainty of the decision, and the reversibility of an incorrect action, will affect the required oversight and control mechanisms. This follows the safety assurance principle that acceptable safety must be justified for a defined application and operating environment, rather than assumed from generic system capability \cite{hawkins_guidance_2022}.

\subsubsection{Current and Future Skill States}
Integrating AI into existing applications may fundamentally change the way that those applications and contexts work. This may be appropriate where the system has been deliberately redesigned around a justified allocation of functions between humans and AI, rather than inserting AI into an unchanged workflow.

However, this does pose an interesting question in the use of operator or SME expertise in evaluating the outputs of a system. For a sufficiently “revolutionary” AI system, would the change to the way the system works invalidate or decrease the expertise of those who were previously experts in conventional systems? How transferrable is this domain expertise, and how can we tell when we have reached the limit of someone’s knowledge \cite{kahn_ai_2024, salgado-criado_human_2025}? For example, someone who is expert in evaluating and cohering intelligence outputs from human analysts may not have adequate understanding of the failure modes of Large Language Models to understand the failure modes unique to \cite{vinay_failure_2025}, and limitations of their use \cite{shafee_evaluation_2025} in processing open-source intelligence information. Clear differentiation between domain SMEs, AI/ML SMEs, safety/security/human factors SMEs, legal advisors, and authorised decision-makers will be required to effectively discuss the complexities of this challenge.

\section{Operational Design Domains/Operational Environment}
\begin{notebox}[Requirements from JSP 936]
	“Where a civil-sector RAS containing AI is procured for Defence use, the military delta in the Operational Design Domain must be identified” (P.36)\\
	“The ODD for the AI should be identified and include information about its context of use and the digital systems in which the AI is designed to operate” (P.38)\\
	“The operating context for the AI components must be clearly defined and communicated to relevant stakeholders” (P.75)\\
	“Appropriate response to reasonably expected inputs outside of the intended design must be defined and demonstrated” (P.78)\\
	“The data should be a sufficiently accurate reflection of the real-world application” (P.157)\\
	“Data should be demonstrated as correct” (P.157)\\
	“AI Assurance must include assurance of behaviour where excursions from the AI ODD may reasonably be expected to occur” (P.209)
\end{notebox}

\subsection{Defining the Operational Environment}
\begin{quote}
	\textbf{\textit{How do we determine that any definition of the operational environment is sufficiently accurate in describing the real-world operational envelope – for anything sufficiently complex? \\ Identifying elements of the ODD is easy – which bits of it matter? When will changes to parts of the ODD change the way that the system behaves?\\Some aspects of the “military delta” will be obvious – but how can we identify if we have captured all aspects of it?}}
\end{quote}
ODDs are often specified using human-centred abstractions of the environment \cite{mehlhorn_ruling_2023}, despite evidence that many AI systems (especially learned perception systems) rely on representations that diverge from human perception \cite{zhang_visual_2018}. Humans intuitively understand variations and expected limitations in their environment, and our assumptions about what information is required to represent the complexity of the real world are not necessarily that which are required by AI systems to construct a machine world model \cite{pramod_computational_2016}. Humans also tend to anthropomorphise machine behaviours, often leading to assuming that AI system’s behaviours represent higher levels of understanding that they actually do - it’s not uncommon for sophisticated combinations of Skills and Rules to imitate higher-level Knowledge and Reasoning understanding modes \cite{cummings_informing_2018}, without the AI system undertaking the actual behaviour represented by a human performing at these levels \cite{curseu_beyond_2026, salles_anthropomorphism_2020}.

Generally, the purpose of an ODD is to seek to define the operational boundaries within which a system is expected to function safely and effectively - in effect creating a set of assumptions and requirements that should drive the realisation of the system \cite{weiss_approach_2024}. However, its purpose needs to extend beyond static definition, in order to capture representative real-world limitations. It must address how we identify gaps, manage uncertainty, and anticipate hazards that arise when assumptions fail.

Managing change throughout the system lifecycle is central to an ODD’s robustness. Real-world conditions evolve, introducing temporary exceptions or permanent drifts that challenge prior assumptions about the capabilities of the system to meet the challenges of these environments \cite{shakeri_formalization_2024}. Similarly, model capabilities shift through retraining and performance updates, creating a dynamic interaction between environment and system competence. ODDs often represent the world as previously experienced, yet this approach falters in the face of edge cases (sometimes called "black swan events"): rare, high-impact scenarios outside historical data \cite{lee_black_2025, ryerson_safety_2020}. Should the system’s ability to perceive and respond to novel phenomena become an explicit component of an ODD? This question indicates the need to integrate capability modelling into domain definition.

Sensor limitations further complicate the picture. What a system can perceive directly influences its operational safety. One illustrative case comes from October 2023, when a Cruise autonomous vehicle failed to detect a pedestrian beneath it and dragged her approximately 20 feet \cite{koopman_lessons_2024}. The system “was unable to detect the pedestrian being dragged… even though [she was] partially in view", indicating limitations in detection, tracking, and scenario handling within the system’s effective ODD. Therefore, an ODD must incorporate sensor constraints and their real-world implications, ensuring that environmental perception aligns with operational expectations.

\subsubsection{Defining an Operational Design Domain}
Consider, for example, a 4-dimensional breakdown, as is sometimes adopted in structured autonomous-vehicle ODD breakdowns \cite{koopman_etal_2019}. Initially, definition is required of the Operational Environment: perhaps including factors such as location, infrastructure characteristics, or communication methods. This grows increasingly complex as the “rules” that govern the environment become more complex – such as a military technology deployed in the chaotic environment of a contested military operating environment. Additional complexity also rears its head when you consider the virtual and social aspects that feature in every modern system: information architectures, system interfaces, human-machine interfaces, political and geopolitical trends, and social and cultural values. 

Add to this complexity, then, the Event Detection and Response characteristics – the system’s ability to perceive that in the world around it. Key here is the detection of all relevant entities – those that matter to the system, as well as management of false positives and false negatives. As well as detecting entities themselves, behaviour must also be determined – what is the expected behaviour of these entities in this environment?  Combine this with the Active System Behaviour and Decisions: the system’s ability to effect change in its environment, such as the set of possible operational outcomes, goal setting and goal seeking behaviours, operational modes and mode transitions. 

Finally, we must also consider the Failure Modes and Fault Management Parameters. These separate into 3 general categories: System Limitation (inherent boundaries or constraints that affect a system's capabilities or performance - not the result of a failure or fault with the system, but rather built in by design or component choice); System Fault (a fault, error, or malfunction in the system such that the system is unable to effectively perform its function); and Fault Responses (aspects of a system-level view of fault detection and mitigation that mitigate system faults).

Importantly, these are not independent characteristics \cite{koopman_etal_2019, koopman_redefining_2024}. They are highly integrated and interdependent on each other. For example, the Event Detection and Response characteristics may well be highly dependent on certain operational modes or mode transitions within Active System Behaviour and Decisions, as the system overheats and processing capabilities shut down (a System Fault from Failure Modes and Fault Management Parameters), leading to a failure to determine all relevant entities in the Operational Environment. 

\subsubsection{Managing the Military Delta}
There are two main aspects to this gap. First, a “military delta” arises from the differing operational requirements of defence systems compared to civilian applications, including environmental conditions, platform integration, and security constraints. Civilian and military technologies evolve along distinct trajectories shaped by these contextual factors \cite{guarascio_digital_2025}, and commercially developed technologies typically require adaptation before they can deliver value in military use \cite{tobias_aebi_unlocking_2025}.

Second, a “military delta” arises from institutional factors, including procurement mechanisms, requirements definition, and standards. Defence acquisition systems employ distinct processes and constraints compared to civilian markets \cite{oishee_kundu_public_2021}, and the transition of commercial technologies into service is heavily mediated by requirements, acquisition, and budgeting processes that are often complex and non-linear \cite{brodi_kotila_fostering_2023}.

A further challenge, beyond identifying an initial set of requirements, is determining - much like in many other areas of AI assurance - if you have captured all of the relevant requirements. Given challenging properties of many AI-enabled systems (e.g. reliance on representative training data, autonomy of real-world actions, human-AI interaction in complex Command and Control systems), AI-enabled systems are likely to increase the difficulty of identification of all relevant requirements. These unarticulated assumptions and tacit requirements (the "unknown-unknowns") are where risk is likely to sit \cite{hillen_navigating_2025, sawyer_unknown_2011} in the adaption and trajectory of adoption of civilian-designed technologies for the military operating environment \cite{guarascio_digital_2025}.

\subsection{Use of the System}
\begin{quote}
\textbf{\textit{How do we cover the usage of the system outside the intended ODD? How can we detect that operation outside the ODD is occurring?\\Is analysing misuse, abuse, and disuse of the system sufficient, or do you need more?}}
\end{quote}
When we define the boundaries of our ODD, we are implicitly setting out the operational space that we expect our system to be used in. However, there are two critically important areas of this space:
\begin{itemize}
	\item reaching the edge of this boundary - where you are “operating under boundary conditions”,
	\item over the edge of this boundary - operating in a space beyond the design.
\end{itemize}
Where this can be particularly challenging is where the boundary is very jagged - i.e. it is not a smooth surface of capability, but rather varies significantly between tasks, actions, or context’s that a human operator would consider to be very similar. This is behaviour exhibited, for example, by frontier Large Language Models (e.g. able to develop complex mathematical equations but can make errors with the date that the work was completed). In practice, inappropriate use of automation outside its defined ODD is a recognised source of safety risk, and has been directly linked to system failures in deployed driving automation systems \cite{cho_operational_2020}.

A second issue is to reliably determine when the system is operating near to or outside of the ODD boundary \cite{shakeri_formalization_2024}. Assuming the ODD is sufficiently broad (e.g. in a general-purpose tool), this becomes a non-trivial problem \cite{hodge_out--distribution_2025}. However, it is important to be able to have confidence in applied guardrails or controls that are used to reduce the risk of system failures. In some situations, it may be a trivial exercise to detect some deviations of an ODD. For example, in a self-driving car where the ODD is primarily based on location, GPS tracking may provide strong indication of likely breach. However, many ODDs represent much more complex behavioural states, where there is not any definitive data that can be linearly mapped to an ODD.

Finally, there is the additional lens of intent to add to the complexity. This contrasts the intent of the system’s designer against that of a human user, and may be summarised through a ‘use, misuse, abuse, disuse’ framework \cite{parasuraman_humans_1997}. For the system designer or analyst, this presents the question: “Is the human user intending to use the output of the AI system in the way the designer intended it to be used?” This addresses some of the principal concerns of highly-autonomous AI-enabled systems – that they are easily used beyond their intended scope of design, and this is where significant failure can occur. Additionally, anthropomorphic interaction can lead users to over‑trust outputs and apply them beyond intended scope\cite{maeda_when_2024}. It can make a meaningful difference to intentionally use the system beyond the ODD (abuse) rather than non-intentionally (misuse). However, this can be seen to apply an overly simplistic lens to a complex problem of human behaviour. For behaviourally complex situations (for example, when operators are under extreme stress), are additional lenses needed?

\section{Dealing with AI as a System of Systems Component}
\begin{notebox}[Requirements from JSP 936]
	“The level of influence and consequences of AI outputs on overall digital system performance should be identified and incorporated into overall system risk analysis” (P.43)\\
	“Where AI interacts with other systems (in particular where they also include AI), the AI behaviour should be understood in the system of systems context” (P.65)\\
	“Analysis for potential effects of reasonable failure modes must be carried out, and must consider the potential for inter-system emergent effects and cascaded failures” (P.82)
\end{notebox}
\subsection{Impact of AI within a system of systems}
\begin{quote}
	\textbf{\textit{How do we “trace” the impact of AI outputs, assumptions, and failure modes throughout a wider system of systems? If the end-user interfaces with a downstream system – how can they understand the dependencies that sit throughout the information/decision chain?}}
\end{quote}
Managing AI within a System of Autonomous Systems (SoAS) introduces significant complexity due to the relationship between autonomy, interdependencies, and emergent behaviours, despite the fact that constituent systems may be operationally and managerially independent \cite{maier_architecting_1998}. The SoAS case is treated here as a high-complexity subset of the broader system-of-systems problem. Each constituent system may operate to its own objectives, with only local observations, while still contributing to a higher-level capability \cite{altmann_emergence_2024}. If the analysis needs to understand the “who/what/when/why/where” for decisions across the SoAS, we need traceability and provenance \cite{mora-cantallops_traceability_2021}. This should not mean exposing every model internal or every intermediate signal from every constituent system – this would just overwhelm operators or those trying to understand the information, and would fail to deliver the true goal of operational explainability \cite{yu_towards_2023, chen_situation_2018}. There is a need to design SoAS to create shared situational awareness and scalable observability: enough context to understand dependencies and consequences, without overwhelming operators with un-interpretable data.

Independent changes in one constituent system can create emergent SoAS-level effects, including behaviours that were not visible when the constituent systems were assessed in isolation, complicating assurance, governance, and risk management \cite{hochmann_designing_2023, altmann_emergence_2024}. When autonomous systems can perceive, predict, plan, and act without real-time human input (in other words, when they have causal agency), this creates challenges in aligning diverse system perspectives, integrating capabilities, and maintaining control when managerial authority is distributed across multiple organizations \cite{macrae_learning_2022}. Distributed managerial authority, conflicting objectives, and unclear accountability can further degrade coordination and decision-making \cite{maier_architecting_1998}. 

AI and autonomy can amplify these challenges by adding opaque decision processes \cite{yu_towards_2023}, data and model dependencies, and new security vulnerabilities \cite{standen_adversarial_2025}. Black-box models can hinder explainability, trust calibration, and V\&V, especially where users cannot relate outputs back to data, assumptions, model behaviour, or intended use \cite{mora-cantallops_traceability_2021}. Additional emergent behaviours or failure modes like miscoordination, conflict, or collusion can also arise across interacting AI component systems \cite{hammond_multi-agent_2025}. These dependencies may be implicit and not documented, including shared training data, common preprocessing assumptions, incompatible reward functions, or undocumented optimisation objectives, reducing visibility and increasing brittleness \cite{mora-cantallops_traceability_2021, altmann_emergence_2024}. 

Testing and assurance become more difficult where AI components evolve through iterative development, tuning, retraining, or configuration change, and where their effects depend on other constituent systems \cite{mora-cantallops_traceability_2021, maier_architecting_1998}.  Adversarial risks are also harder to bound in multi-agent settings because effects can propagate across time and between agents \cite{standen_adversarial_2025}. These dynamics can create brittle SoAS behaviour, where small perturbations, network effects, or feedback loops create disproportionate SoAS-level consequences \cite{hammond_multi-agent_2025}.

SoASs containing interacting AI agents may be vulnerable to adversarial exploitation, particularly where attackers can perturb observations, communications, or learned policies, and so can take advantage of these vulnerabilities and instabilities to create cascading failures \cite{standen_adversarial_2025}. These failures can be hard to prevent and hard to recover from because the triggering condition may be distributed across agents, time, data, and interfaces \cite{standen_adversarial_2025, macrae_learning_2022}. Not only do they often require significant rework to re-establish these complex systems, but the nature of these complex causal-chain failures often leads to additional difficulties in reconstructing evidence, tracing causal chains, and assigning accountability across technical, organisational, and contractual boundaries \cite{macrae_learning_2022}. 

Finally, there is a practical assurance challenge in deciding which AI-influenced decisions matter at SoAS level. Autonomous systems may make frequent decisions at different levels of abstraction, from low-level filtering and routing to goal- or mission-level planning. When AI models are underpinning the decisions of autonomous systems, it creates increasing difficulty in identifying which of these decisions are consequential to the behaviour of the system overall. For example, changes to preprocessing, filtering, observation functions, reward weights, fusion logic, or threshold settings may alter the behaviour of a constituent system and produce SoAS-level effects \cite{mora-cantallops_traceability_2021, altmann_emergence_2024}. These microdecisions can be high-frequency, context-sensitive, opaque to users, and difficult to reproduce where randomness, retraining, or configuration changes affect the pipeline. The assurance problem is therefore to identify which AI-influenced decisions are safety-, mission-, or accountability-significant at SoAS level. 

\subsection{Interfaces with existing and conventional systems}
\begin{quote}
	\textbf{\textit{When many constituent systems are conventional systems, may already exist, and may have been developed under different assurance regimes, how do we determine the effect of an AI system on them and on the outputs they produce?}}
\end{quote}

A significant integration risk is the interface between new AI-enabled components and external legacy systems \cite{terry_systems_2025}. This presents a number of challenges for assurance, above and beyond that of purely assuring the new AI system alone:
\begin{itemize}
	\item Interfaces often encode assumptions about determinism, timing, data quality, stability, and failure behaviour that may not be fully captured in Interface Control Documents. Assurance should therefore test conformance to the intended interface behaviour, not only conformance to the documented interface \cite{maier_architecting_1998, terry_systems_2025}.
	\item Adding AI-enabled components to a legacy SoS can change the attack surface of both the AI component and the wider SoS. Security assurance should therefore consider adversarial effects that propagate through interfaces, not only attacks against the new AI component itself \cite{standen_adversarial_2025}.
	\item Iterative AI development, including data changes, retraining, model updates, and parameter tuning, is difficult to reconcile with legacy systems that rely on static assurance evidence and fixed baselines \cite{mora-cantallops_traceability_2021, terry_systems_2025}.
\end{itemize}

\section{Assessing and Maintaining AI Performance}
\begin{notebox}[Requirements from JSP 936]
	“When the system has more than one mode of operation or level of autonomy, the impact analysis must be conducted for all modes and autonomy levels” (P.54)\\
	“Analysis for potential effects of reasonable failure modes must be carried out” (P.82)\\
	“For collaboratively trained human-AI teams, assurance of the overall behaviour for each team must be provided on a case-by-case basis” (P.131)\\
	“Where wider system requirements include the expectation that AI will be modified in-situ then planning should include how this will be controlled and assured for continuing confidence” (P.135)\\
	“Performance requirements for the AI should be clearly stated alongside functional requirements” (P.140)\\
	“The AI architecture must be traceable to requirements and be able to incorporate the intended behaviours whilst protecting against entry into failure modes identified during hazard analysis” (P.150)\\
	“Safe behaviours of the AI when exposed to inputs that fall outside of the ODD must be demonstrated” (P.171)
\end{notebox}
\subsection{Instance Specific AI Performance Assessment}
\begin{quote}
	\textbf{\textit{Is assurance at each level of autonomy required or practical? What can be reused?\\
			How do you determine what reuse of assurance is acceptable, and what is no longer true?\\
			The same applies to the assurance of overall behaviour for every team – in order to practically carry this out, is the level of assurance going to provide sufficient mitigation?\\
			What change is required to the team (either human or AI) before re-assurance is required?
	}}
\end{quote}
Levels of autonomy should not be treated as a single linear scale for assurance purposes \cite{parasuraman_model_2000}. Where a system can operate in different modes or at different autonomy levels, the assurance argument should clearly demarcate between different operating modes or states \cite{yang_framework_2024, eom_mode_2022}. This is necessary because automation changes the operator's role\cite{bainbridge_ironies_1983}, can create out-of-the-loop performance problems \cite{endsley_out-loop_1995}, and can introduce mode-transition hazards \cite{yang_framework_2024, methnani_let_2021}. Clearer mode definitions also provide a stronger basis for resilience testing and graceful degradation \cite{weiss_approach_2024, hawkins_guidance_2022}.

Human-AI teaming adds a further assurance challenge because the relevant behaviour is not only the AI model's behaviour, but the joint behaviour of the human-AI team \cite{kudina_sociotechnical_2024, gonzalez_toward_2026}. Human reliance can change over time through a wide range of factors, including trust calibration, over-trust, under-trust, workload, situation awareness, and automation bias \cite{wischnewski_measuring_2023, parasuraman_complacency_2010, lee_trust_2004, hoff_trust_2015, okamura_adaptive_2020}. If the AI also adapts to the preferences, decisions, or operating patterns of a specific user, the assured object is no longer only a fixed model or generic role allocation, but a changing socio-technical pairing \cite{ashmore_assuring_2022, asaadi_dynamic_2020}. This also creates a system-of-systems assurance problem \cite{maier_architecting_1998, kudina_sociotechnical_2024}. Multiple nominally identical human-AI teams may diverge due to this adaptive or evolutionary model \cite{okamura_adaptive_2020}. The practical question is therefore not whether assurance must be repeated from first principles for every instance, but which claims, assumptions, and evidence remain valid across instances \cite{cheng_ac-ros_2020}, how individual assurance of these instances is possible, and whether the whole system can maintain justification for assurance given this adaptive change \cite{asaadi_dynamic_2020, bloomfield_assurance_2021}.

\subsection{Management of AI Performance}
\begin{quote}
	\textbf{\textit{How should AI-component-specific performance requirements trace to system-level performance and mission outcomes?
	}}
\end{quote}
AI-specific performance requirements can be useful, but need to be traceable to system-level outcomes. Model metrics (e.g. accuracy, precision, recall, F1-score) may be necessary evidence, but they should not be treated as substitutes for system performance requirements \cite{thieu_permetrics_2024, habibullah_non-functional_2023}. It’s important to be clear that the priority needs to be measures of system outcome, and then AI model performance outputs should support achieving those outcomes \cite{koopman_etal_2019, weiss_approach_2024}. Otherwise, there is a risk of optimising the model against metrics that do not adequately represent operational success, safety, or human-AI team performance \cite{ashmore_assuring_2022, lever_facing_2025}.

\subsection{Defining and Understanding AI Failure Modes}
\begin{quote}
	\textbf{\textit{If the failure mode lies within the AI’s internal ‘black box,’ how can you determine the point of entry or credible initiating causes where the failure mode originates?\\
			How can we define and manage the functional interactions between human and AI? How can we understand the basis for those interactions, as well as how they can fail?\\
			What is meant by `safe behaviour': a locally acceptable model output, a safe system action, or a safe system outcome under defined operating conditions?
	}}
\end{quote}
Analysing AI failure modes is difficult because failures may arise across the entire scope of the deployed socio-technical system \cite{kumar_failure_2019, ashmore_assuring_2022}. Some specific challenges include:
\begin{itemize}
	\item Failures are often interdependent. A failure on one side of the human-machine interface can change behaviour on the other side, creating feedback effects that propagate through the wider system \cite{parasuraman_humans_1997, jimeno_altelarrea_stpa_2022}.
	\item Conventional techniques such as FTA and FMEA remain useful, but they are insufficient on their own for AI-enabled systems \cite{jimeno_altelarrea_stpa_2022, rismani_silos_2024}. They work best where component functions, failure causes, and causal paths are relatively well defined. AI-enabled systems add new failure mechanisms, often drawn from a broad and widespread set of origins that creates a failure surface sitting beyond the scope of conventional failure analysis techniques \cite{quinonero-candela_dataset_2008, czarnecki_steam_2023, jimeno_altelarrea_stpa_2022, hawkins_guidance_2022}. A multi-method analysis is therefore required. The analysis should combine model-focused, data-focused, human-factors, interaction, operational-context, and system-safety perspectives \cite{hawkins_guidance_2022, yang_framework_2024, rismani_silos_2024}. Each perspective is incomplete on its own, but together they provide a more defensible failure analysis \cite{bloomfield_assurance_2021}.
	\item AI integration can increase the distance between initiating cause and observed failure. This can be due to the increased failure surface, as described previously \cite{ashmore_assuring_2022, weiss_approach_2024}. The observed failure may therefore appear late in the causal chain, after several technical and human decisions have interacted \cite{koopman_lessons_2024, kudina_sociotechnical_2024}.
	\item AI-enabled systems also increase the need to analyse human-cyber-physical failure modes. Many conventional systems already require this, but AI systems often depend on both tight coupling across human-cyber-physical interactions \cite{endsley_supporting_2023, sterz_quest_2024}, combined with an increase brittleness in human-machine interactions \cite{methnani_let_2021, marusich_trust_2025}, and this can lead to a significant diversification in failure modes as well as limited ability to recover from failure.
\end{itemize}
Failure analysis should distinguish model-performance failures from system-outcome failures. A model can satisfy its local performance metric while the system still fails to achieve the intended operational, mission, or safety outcome \cite{rismani_silos_2024}. Conversely, a model error may be tolerated if the system architecture, human oversight, fallback behaviour, or degraded mode prevents harm. Model metrics should therefore be treated as evidence within a system safety argument, not as the definition of safety \cite{ashmore_assuring_2022, hawkins_guidance_2022, kumar_failure_2019}

\section{Analysing Safety and Security in AI-Enabled Systems}
\begin{notebox}[Requirements from JSP 936]
	“The AI design, within the context of the system in which it operates, should minimise insofar as is reasonably practicable the adversarial attack surface” (P.81)\\
	“A Hazard Analysis must be undertaken to identify hazards introduced through the use of AI” (P.141)\\
	“AI may have unique safety risks associated with its development or behaviour. These must be analysed and included in the relevant wider safety cases and software and system risk assessments” (P.191)
\end{notebox}
\subsection{Integration of Safety and Security}
\begin{quote}
	\textbf{\textit{Given existing safety and security processes are not particularly integrated, how do we manage this in the AI space where it is even more crucial?
	}}
\end{quote}
Systems engineering, safety engineering, and security engineering within Defence are often coordinated at governance or acquisition level, but less well integrated at the development level, where requirements, architecture, hazards, threats, controls, and evidence are produced \cite{macher_architectural_2021, ross_engineering_2022}. For AI-enabled systems, this separation is more consequential because security, safety, performance, and trustworthiness claims are interdependent \cite{vassilev_adversarial_2024, lin_ai_2025}. For example, a prompt injection attack against a large language model can affect not only confidentiality or integrity, but also system safety and mission outcome \cite{gulyamov_prompt_2026}. The barrier is not only cultural. Different disciplines often use different modelling tools, analysis methods, terminology, standards, and evidence repositories. As a result, common artefacts (like hazards, assumptions, and risks) may be analysed separately even when they concern the same system behaviour \cite{ross_engineering_2022, rismani_silos_2024}. Secure systems engineering approaches, such as NIST SP 800-160 Vol. 1 for engineering trustworthy secure systems \cite{ross_engineering_2022}, provide a useful starting point. However, for AI-enabled systems they should be extended across the system lifecycle, to ensure that attacker goals, objectives, and capabilities are integrated into system design considerations \cite{vassilev_adversarial_2024}.

\subsection{Additional Safety and Security Vulnerabilities of AI System}
\begin{quote}
	\textbf{\textit{Which vulnerabilities are specific to, amplified by, or made harder to detect by AI-enabled systems compared with conventional software-intensive systems?
	}}
\end{quote}
AI-enabled systems retain conventional engineered system security objectives, including confidentiality, integrity, availability, authenticity, accountability, and non-repudiation. The difference lies in the mechanisms by which threats to these objectives may exist: AI introduces additional assets and attack paths (e.g. training data, validation data, labels, model weights, prompts, embeddings, tool interfaces, model APIs, feedback loops, and model provenance \cite{sidhpurwala_building_2025}). Standard threat modelling approaches such as STRIDE remain useful, but they should not be treated as complete for AI-enabled systems. They may need to be combined with privacy threat modelling (e.g. LINDDUN), adversarial ML taxonomies, AI lifecycle analysis, supply-chain analysis, misuse-case analysis, and hazard analysis \cite{vassilev_adversarial_2024, rismani_silos_2024, sidhpurwala_building_2025}.

AI-enabled systems also introduce or amplify threat categories that are not well covered by conventional software security analysis. Adversarial attacks (such as data poisoning, evasion attacks, or model extraction) can cripple models \cite{vassilev_adversarial_2024, kumar_failure_2019}. However, another contributing factor is that some attacks may be difficult to detect because the model can continue to appear functional while its behaviour is selectively degraded, biased, exfiltrating information, or unsafe under specific triggers or operating conditions \cite{vassilev_adversarial_2024, sidhpurwala_building_2025}.

Certain classes of models also have individual vulnerabilities, or may be affected in different ways by attack classes. For example, LLM-based systems are vulnerable to prompt injection and indirect instruction attacks, while agentic and multi-agent LLM systems add risks through tool use, memory, inter-agent communication, delegated authority, and external data retrieval \cite{gulyamov_prompt_2026, he_red-teaming_2025}. Conformity bias (the behavioural feature of LLMs that move towards conformity, rather than diversity) can lead to abnormalities not being highlighted or low-magnitude signals being amplified, or monoculture vulnerabilities (where multiple agents are created from the same base model) can give lead to inherent behaviour patterns interacting to create a vulnerability \cite{reid_risk_2025}. 

The use of third-party models, externally hosted APIs, open-source model weights, model repositories, plugins, datasets, and orchestration frameworks creates an AI supply-chain risk \cite{vassilev_adversarial_2024, sidhpurwala_building_2025}. Building a dependable AI system cannot only include cyber assurance of directly created AI systems, but also on the foundations on which those systems rely \cite{ross_engineering_2022, ashmore_assuring_2022}.  

\section{Measuring Ethicality}
\begin{notebox}[Requirements from JSP 936]
	“In cases where an AI system presents unacceptable negative ethical risks … deployment or development must be halted in a safe manner until risks can be sufficiently managed” (P.91)\\
	“Defence must behave ethically” (P.95)\\
	“Defence must be seen to be ethical” (P.95)
\end{notebox}
\subsection{Subjectivity of Defence}
\begin{quote}
	\textbf{\textit{Ethical judgements are often subjective, value-laden and context-dependent. How do we capture this subjectivity in objective reporting?\\
	How do we benchmark and present ethical risks, even when ethics are subjective, and different stakeholders have different views on acceptability?
	}}
\end{quote}
Ethical risk appetite is variable and context-dependent. It may differ between stakeholders, roles, operational settings, and levels of organisational accountability \cite{kallina_stakeholder_2024, krijger_enter_2022}.  The variety in contextual considerations means that defining fixed guardrails is very challenging, and unlikely to be suitable unless the ethical boundaries are very clear \cite{canca_operationalizing_2020}. In a military context, Article 36 weapons review in a Defence context adds a further cross-disciplinary interface. Technical teams, operational warfighters, legal advisers, commanders, and policy authorities may each hold evidence needed to judge whether a new weapon, means, or method of warfare, especially when a weapon system is capable of continuing to ‘learn’ on its own after being deployed on the battlefield, is lawful and acceptable in its intended context of use \cite{mcfarland_legal_2023}.

Quantifying ethical harms is difficult because many relevant harms are socio-technical rather than purely technical \cite{selbst_fairness_2019,stahl_systematic_2023}. Metrics for fairness, bias, collateral effects, explainability, proportionality, or unacceptable harm are be immature, contested, context-dependent, and can also require trade-offs due to mutual incompatibility \cite{krijger_enter_2022, canca_operationalizing_2020}. 
A further challenge lies in interpreting what constitutes desirable versus undesirable behaviour within these metrics. In particular, bias is not inherently negative: in machine learning, some form of inductive bias is necessary for any system to generalise beyond observed data and therefore to make decisions at all. An entirely inductively unbiased system would be unable to prioritise outcomes or act under uncertainty \cite{wolpert_no_1997, mitchell_need_1980}. The concern is therefore not the presence of bias per se, but whether that bias is appropriate to the task and aligned with acceptable objectives, or whether it creates socially harmful biases and outcomes \cite{goyal_inductive_2022}.
Unintended, unexamined, or misaligned biases, however, can lead to discriminatory or otherwise harmful outcomes. These forms of bias are ethically problematic and can also undermine system performance and trust, creating both normative and practical risks \cite{selbst_fairness_2019, krijger_enter_2022, ferrara_fairness_2023}.

Ethical acceptability is also not limited to internal assessment. Defence must be able to justify both the substance of the decision and the process by which it was reached \cite{stahl_systematic_2023, blanchard_ethical_2025}. Public confidence, parliamentary scrutiny, media scrutiny, alliance expectations, and perceived legitimacy can affect whether an AI-enabled capability is acceptable in practice, even where a narrow technical or legal assessment is satisfied \cite{jobin_global_2019, ai_in_weapon_systems_committee_proceed_2023}.
\subsection{Ethicality as Part of Defence}
\begin{quote}
	\textbf{\textit{If one part of the Defence enterprise is unethical, does this compromise the ethicality of Defence as a whole?
	}}
\end{quote}
Individual projects cannot be responsible for the ethical standing of the whole Defence enterprise. They can, however, create ethical risk for the wider enterprise where their outputs, data, models, decisions, suppliers, or operational effects are reused across systems of systems \cite{maier_architecting_1998,kudina_sociotechnical_2024}.
Ethical assessment should therefore cover the individual AI-enabled system, the operational system into which it is integrated, and the wider system-of-systems effects created by its use \cite{kudina_sociotechnical_2024}. The assessment should consider how the AI system’s autonomy, scale, speed, opacity, data dependencies, and human oversight arrangements change the ethical risk profile of the wider capability \cite{selbst_fairness_2019, sterz_quest_2024}.

Ethical assessment also needs to be lifecycle-based. It is not enough to assess ethical acceptability at integration or deployment \cite{ashmore_assuring_2022,stahl_systematic_2023}. Ethical-by-design activities should begin at concept and continue through requirements, architecture, data selection, model development, verification, operational monitoring, update control, and withdrawal \cite{canca_operationalizing_2020,ashmore_assuring_2022}. This reduces the risk that unacceptable ethical risks emerge late, when design changes are more expensive and assurance evidence is harder to reconstruct \cite{stahl_systematic_2023,canca_operationalizing_2020}.

\section{Mitigating the Inherent Complexities of AI}
\begin{notebox}[Requirements from JSP 936]
	“All models should be transparent” (P.167)\\
	“All models should include appropriate explanations of their output” (P.167)\\
	“All models should provide measures of uncertainty that are understandable to the various stakeholders” (P.167)
\end{notebox}
\subsection{Inherent Complexities}
\begin{quote}
	\textbf{\textit{How do we address failures and weaknesses in AI systems where the nature of the systems themselves is essential to the failure of the system?
	}}
\end{quote}

Several requirements point to areas where current methods provide potentially useful evidence but not complete assurance:
\begin{itemize}
	\item \textbf{Model transparency:} Some AI model classes, especially complex deep learning systems, lack interpretable internal representations, and can block causal mapping \cite{ribeiro_why_2016, rudin_stop_2019}. Many post-hoc explanation methods are approximations, not guarantees of correctness, further limiting their usefulness for assurance evidence \cite{tomsett_rapid_2020, ashmore_assuring_2022}.
	\item \textbf{Appropriate explanations:} Explanations need to be faithful enough to support the assurance claim and understandable enough for the stakeholder using them \cite{tomsett_rapid_2020}, but many explainable AI (XAI) methods trade fidelity for simplicity \cite{lipton_mythos_2018}. Careful management of explanations is also required to support trust calibration - misleading or subjective explanations could impact perceived system capability inaccurately \cite{okamura_adaptive_2020, tomsett_rapid_2020, wischnewski_measuring_2023}.
	\item \textbf{Measures of uncertainty:} relying on uncertainty measures from AI systems is challenging \cite{adel_trustworthy_2024}. Common confidence scores may be poorly calibrated and may not distinguish aleatoric uncertainty, which arises from inherent variability or noise, from epistemic uncertainty, which arises from limited knowledge or insufficient data \cite{guo_calibration_2017, wimmer_quantifying_2023}. Bayesian methods, ensembles, conformal methods, and sampling approaches can improve uncertainty estimation, but they add computational cost, modelling assumptions, and integration complexity \cite{wimmer_quantifying_2023, tomsett_rapid_2020}. In a system-of-systems context, uncertainty can also propagate and compound through decision chains and information streams \cite{hochmann_designing_2023}.
	\item \textbf{Non-deterministic behaviour:} Learning systems produce variable outputs under identical inputs due to factors including stochastic training, stochastic inference, or online learning. Every update may therefore invalidate previous claims, assumptions, arguments, or evidence \cite{asaadi_dynamic_2020, amershi_software_2019}.
	In multi-agent systems, interactions between agents can create emergent behaviours that are not evident from testing each agent in isolation \cite{hochmann_designing_2023,he_red-teaming_2025}.
	\item \textbf{Brittleness and robustness:} AI models can often fail under distribution shift, or in extremes of training data (black swan events) \cite{quinonero-candela_dataset_2008, hodge_out--distribution_2025, weiss_approach_2024}. Robustness metrics are useful, but they should not be treated as proof of real-world robustness unless they are linked to the intended operating context and assurance case \cite{hodge_out--distribution_2025, ashmore_assuring_2022}. Brittleness also affects trust calibration: unexpected failures can cause over-correction, under-reliance, or loss of confidence even where the system remains useful within its defined limits \cite{tomsett_rapid_2020, okamura_adaptive_2020}.
\end{itemize}
These issues concern model outputs, but assurance must ultimately address system outcomes. Transparency, explanations, uncertainty estimates, robustness tests, and non-determinism controls are evidence for the safety and acceptability of the system \cite{ashmore_assuring_2022,hawkins_guidance_2022}. They are not, on their own, proof that the system will achieve the intended operational, safety, legal, or ethical outcome \cite{habibullah_non-functional_2023,ashmore_assuring_2022}. They should be treated as evidence within an explicit assurance argument \cite{habli_big_2025}.

\section{Conclusion}
This report identified eight recurring challenge areas in the operationalisation of JSP 936 Part 1: evidence adequacy, human-AI interaction, operational environment definition, system-of-systems integration, AI performance management, safety and security analysis, ethical assessment, and the inherent complexity of AI-enabled systems. The analysis does not suggest that JSP 936 is deficient as a directive. Rather, it shows that implementation depends on translating policy requirements into defensible technical, operational, and organisational assurance practice.

A recurring finding is that AI assurance cannot be reduced to model testing, documentation, impact assessment, or governance review in isolation. Confidence depends on the relationship between evidence, assumptions, hazards, human roles, operating conditions, ethical considerations, security threats, and system-level outcomes. Evidence that is adequate for one system boundary, autonomy level, operational design domain, or human-AI team may not remain adequate when those conditions change.

AI-enabled Defence systems require assurance methods that are explicit about uncertainty, traceable across system boundaries, sensitive to operational context, and adaptable as systems, data, models, users, and threats change. Further work should focus on practical methods that connect JSP 936 requirements to assurance arguments, evidence models, design decisions, safety and security analysis, ethical review, and operational monitoring. These methods should help teams determine what must be true for justified confidence in an AI-enabled Defence system, and what evidence is sufficient to support that confidence in a defined operational context.


\section*{The Authors}
\paragraph{Callum Cockburn}
	Callum is Senior Technical Innovation Manager at Synoptix, where he leads R\&D programmes focusing on novel or disruptive technologies and working from fundamental research through to commercial product development.

	Callum currently focuses on AI technologies, especially Responsible AI, AI Assurance, Assured Capability and Cyber Resilience, and Human-Autonomy Teaming. He also leads Synoptix’s AI Assurance capability and has driven business-wide AI adoption and governance, including AI policy, training, internal governance and responsible use of AI tools. He is an Incorporated Systems Engineer by background and trained as a Mechanical and Biomedical Engineer.
	
	His research interests include complex data systems, human-autonomy teaming, multi-disciplinary analysis of complex socio-technical systems, process mining and digital twins in safety-critical environments, and high-stakes AI assurance. 
\paragraph{Sam Farrow}
	Sam is Technology Director at Synoptix, and he is an experienced engineering leader who has worked for more than 15 years in the development of cutting-edge technology across the defence, security and critical national infrastructure domains. With a robust background in Systems Engineering and extensive experience in product development, Sam brings a blend of strategic insight and hands-on expertise to the development and integration of technology.
	
	Across a varied career, Sam has successfully led ambitious product development and research projects across novel weapons, combat air systems, uncrewed systems, mission planning, computer vision, cyber security and AI assurance.

\printbibliography

\end{document}